\documentclass[num-refs]{wiley-article}

\usepackage{siunitx}
\usepackage{comment}
\usepackage{float}
\usepackage{caption}
\usepackage{graphicx}
\usepackage{booktabs}
\usepackage{ulem}

\newcommand{\bD}{ {\bf D} }

\newcommand{\be}{\begin{enumerate}}
\newcommand{\ee}{\end{enumerate}}
\newcommand{\beq}{\begin{equation}}
\newcommand{\eeq}{\end{equation}}
\newcommand{\bi}{\begin{itemize}}
\newcommand{\ei}{\end{itemize}}
\newcommand{\beas}{\begin{eqnarray*}}
\newcommand{\eeas}{\end{eqnarray*}}

\newcommand{\bdes}{\begin{description}}
\newcommand{\edes}{\end{description}}

\papertype{Research Article}

\title{Optimizing Interim Analysis Timing\\ for Bayesian Adaptive Commensurate Designs}

\abbrevs{BCI, Bayesian credible interval; EHSS, Effective Historical Sample Size; FDA, Food and Drug Administration; IA, Interim Analysis; PDUFA, Prescription Drug User Fee Act; RWE, Real World Evidence}

\author[1\authfn{1}]{Xiao Wu}
\author[2\authfn{1}]{Yi Xu}
\author[3\authfn{1}]{Bradley P. Carlin}

\contrib[\authfn{1}]{Equally contributing authors.}

\affil[1]{Department of Biostatistics, Harvard T.H.\ Chan School of Public Health, Boston, MA, USA}
\affil[2]{Xenon Pharmaceuticals Inc., Burnaby, BC, Canada}
\affil[3]{Counterpoint Statistical Consulting, LLC, Edina, MN, USA}

\corraddress{Xiao Wu, Department of Biostatistics, Harvard T.H.\ Chan School of Public Health, Boston, MA, USA}

\corremail{wuxiao@g.harvard.edu}

\fundinginfo{The work of all three authors was supported by Sanofi Pharmaceuticals.}

\runningauthor{Wu et al.}

\begin{document}

\maketitle

\begin{abstract}
In developing products for rare diseases, statistical challenges arise due to the limited number of patients available for participation
in drug trials and other clinical research.  Bayesian adaptive clinical trial designs offer the possibility of increased statistical efficiency,  reduced development cost and ethical hazard prevention via their incorporation of evidence from external sources (historical data, expert opinions, and real-world evidence), and
flexibility in the specification of interim looks. 
In this paper, we propose a novel Bayesian adaptive commensurate design that 
borrows adaptively from historical information and also
uses a particular payoff function to optimize the timing of the study's interim analysis. 
The trial payoff is a function of how many samples can be saved via early stopping and the probability of making correct early decisions for either futility or efficacy.  
We calibrate our Bayesian algorithm to have acceptable long-run frequentist properties (Type I error and Power) via simulation at the design stage.
We illustrate our approach using a pediatric trial design setting testing the effect of a new drug for a rare genetic disease. The {\bf optimIA} {\tt R} package available at \url{https://github.com/wxwx1993/Bayesian_IA_Timing} provides an easy-to-use implementation of our approach.

\end{abstract}

\section{Introduction}
The need for more efficient clinical trial methods continues to increase.
Developers of new drugs and medical devices are under increasing pressure to control development costs, especially in the clinical testing phase.  In the U.S., regulators
at the Food and Drug Administration (FDA) have been motivated since December 2016 by the 21st Century Cures Act and corresponding regulatory rule changes in the Prescription Drug User Fee Act (PDUFA) VI.
These documents have encouraged FDA to consider
Phase II and even Phase III applications that utilize novel statistical methods that
borrow from previous clinical data and perhaps even real world evidence (RWE) \cite{wechsler2016pdufa,sobel2018real,jain2019pdufa}.

Bayesian clinical trial designs offer the potential advantages of reduced study sample size, increased statistical power, and reductions in cost and ethical hazard \citep{hobbs2011hierarchical}. 
In this paper, we propose a Bayesian adaptive statistical approach \citep{carlin2008bayesian},
implemented using commensurate priors \citep{hobbs2011hierarchical}, and
utilizing a novel ``payoff function" to select an optimal time to 
perform an interim look at the data.
Our Bayesian adaptive approach gets the most out of available data by
(a) permitting borrowing from adult data in our pediatric setting, and (b)
by allowing the study to terminate early (at the interim look) if the novel treatment emerges as unequivocally better than placebo (``early win"), or fails to deliver some minimum level of efficacy (``futility").
These features allow reduction of total trial duration, thus reduce cost and ethical hazard.
Statistically, adaptive trials are most easily implemented using a {Bayesian} framework
(see e.g. \cite{berry2010bayesian}), since it avoids problems with traditional $p$-values
and ``alpha-spending functions" (reviewed by e.g. \cite{demets1994interim}), instead directly computing
the probability that each treatment is effective given the
available data (a {\it posterior} probability calculation).  Bayesian procedures also more readily permit
incorporation of external evidence (such as historical data and expert opinion) when needed and appropriate.

A specific example motivating this research was the consideration of a pediatric trial design to test the effect of a oral drug for Gaucher disease, a rare genetic disease belonging to the class of lysosomal storage disorders \cite{grabowski2015gaucher}.  In 2014, FDA granted approval for this drug as a first-line treatment for adults with Gaucher disease type 1 who have a CYP2D6 extensive, intermediate, or poor metabolizer phenotype based on two pivotal studies \cite{mistry2015effect,cox2015eliglustat}. 
 In particular, efficacy in treatment-naive patients was demonstrated in the placebo-controlled ENGAGE trial \cite{mistry2015effect}, which enrolled patients with Gaucher disease type 1 who were at least 16 years of age, with the primary endpoint being reduction in spleen volume (percent change from baseline).
In order to extend the label to treatment-naive children (under age 16), a pediatric study was needed.  However, there were significant challenges in conducting an adequately powered placebo-controlled study in the treatment-naive pediatric population, due to very slow expected enrollment, resulting in a high likelihood that the trial will be unable to fully enroll enough patients to achieve acceptable power. 
We would also expect challenges to pediatric patients (especially those assigned to placebo) in remaining compliant with the study protocol.  

The questions raised by this example motivate us to consider an alternative, adaptive commensurate study design to maximize the information available at an interim analysis (IA).  This design cautiously borrows from the adult data when appropriate, and potentially stops the study after enrolling fewer patients without sacrificing statistical validity. In the case of our pediatric study setting, it was reasonable from a clinical perspective to assume that the primary endpoint used to measure the treatment effect in adult populations would still be appropriate for pediatric patients, and that the magnitude of change in the primary endpoint would likely be similar between adults and children in both the placebo and treatment arms. In this situation, common to many pediatric study designs \cite{bavdekar2013pediatric}, the commensurate prior approach for incorporating information from historical data \cite{hobbs2012hierarchical,murray2014semiparametric,viele2014use,van2018including,
lewisetal2019} can be useful . 

The IA is highly desirable in the above example, since the fixed sample size design may require a sample size that is not realistic, with the result that we cannot finish the study in a realistic period of time.  On the other hand, although the Bayesian method allows us to assess posterior probabilities of futility and efficacy continuously as the data accumulate, due to the significant cost involved in cleaning and making the database available for IA, multiple IAs are not desirable.   Therefore, it’s important to determine an optimal time to perform the single IA that provides the maximum chance of making a correct early decision. 
Papers investigating the optimal placement of interim analyses do not appear plentiful in the literature. The most popular optimal design in the clinical area is the Simon two-stage design \citep{simon1989optimal}, which minimizes either the trial's expected or maximum sample size. However, the Simon design restricts to a binary response, and uses binomial probabilities to determine the optimal first- and
second-stage sample sizes to maximize power subject to Type I error constraints.  Early  stopping  is permitted only for futility (not success), and the timing of the IA is also fixed before data collection begins.
Hampson et al. \citep{hampson2014bayesian}
offer a Bayesian design that incorporates external information arising from expert opinion or historical  data.   However, the procedure relies on a subjective expert determination regarding how much historical information to borrow.  In addition, the
design's adaptation is only on the choice
of the allocation ratio between two arms; there is no explicit interim analysis proposed in their design.
Togo and Iwasaki \cite{TogoIwasaki2013} proposed a method that seeks to minimize the total expected sample size under a specified (continuous) treatment effect, and find that, regardless of the effect size, the optimal time for a single IA is at approximately $2/3$ of the planned sample size for the O'Brien-Fleming-type and approximately $1/2$ for the Pocock-type alpha spending functions, where the expected sample sizes were calculated under a fixed treatment effect as used for the study power. They also noted that when the true effect size was better than or worse than the planned treatment effect, the optimal time would be shifted. In practice, the timing of an interim analysis for non-Bayesian type of studies was typically chosen in the range of $40-60 \%$ of the total sample size based on number of patients needed for safety assessment and enrollment estimation to allow potential saving with the early stopping. 
Yet a goal of clinical trials is often to seek to optimize the tradeoff between costs (e.g., the expected sample size) and benefits (e.g., the correct early futility/efficacy decisions at IA), along with many other consideration that go beyond a standard sample size calculation \cite{anderson2014timing}. 

The Bayesian paradigm is especially promising for constructing our adaptive framework, since it provides a unified and interpretable language for data collection, inference, and decision making \cite{parmigiani2002modeling}. However, on the Bayesian side, the literature for investigating optimal IA timing is even sparser.
A rare exception is the work of Huang and Fu \cite{HuangFu2016}, who use simulation to estimate the optimal location of a single IA using a utility-based Bayesian adaptive design in a particular dose-response setting. We hope the new Bayesian designs proposed in this paper can be soon applied in the future clinical development program in the comparable situations. 

The rest of our paper is organized as follows. Section~\ref{method} lays out the details of our Bayesian adaptive commensurate prior approach, along with a step-by-step algorithm for its implementation.
Section~\ref{results} then gives the results of an extensive simulation study to check our method's performance in the pediatric example setting.  Our approach is able to obtain sensible optimal look times that maximize a payoff function that is essentially measures the weighted conditional probability of early stopping relative to the total sample size expected.  
Finally, Section~\ref{discussion} summarizes our findings and offers avenues for future research, including alternate definitions of the payoff function.

\section{Statistical Methods and Algorithmic Approach}
\label{method}

In our approach, 
we apply Bayesian methods with a commensurate prior to potentially stop a study at a single interim analysis.  We use early futility and efficacy criteria based on Bayesian posterior probabilities of a treatment difference reaching pre-specified thresholds \citep{berry2010bayesian}, after adaptively 
borrowing information from historical adult study data, dependent on its similarity to data from the current study. We calibrate our Bayesian procedures to have acceptable long-run frequentist properties (Type I error and power) via computer simulation at the design stage \citep{psioda2018bayesian}.  The optimal timing of the IA will be evaluated via simulation by assessing a grid of plausible time points, among all decision criteria that meet certain Bayesian and frequentist properties. This optimization is based on a payoff function that characterizes the benefit/cost ratio of the decision. The proposed payoff function also introduces a weight parameter that allows expert input, including level of interest and/or confidence on the new treatment, available budget, and the internal and external competitive environment. 

To formalize ideas, let $n_k$ be the sample size for the pediatric study in group $k$, where $k=1$ is the placebo group and $k=2$ is the treatment group. Let $n_{0k}$ be the sample size for the two historical (adult) groups respectively.  Since we will typically have much more adult data than pediatric, in what follows we set
$n_1 = n_2 < n_{01} = n_{02}$. Let  $n \equiv n_1+n_2$ be the maximum total pediatric sample size if the trial runs to completion.  We recommend selecting the maximum pediatric sample size to achieve a reasonable power based on a clinically meaningful target treatment effect, in order to allow the study to still have a good chance to achieve its objective in the least favorable
case where no information at all can be borrowed from the adult study data.  We propose a trial with a single interim look, after $n'$ pediatric spleen reductions have been observed.
Let $\theta_k$ be the mean $\%$ reduction in spleen volume for children in group $k$ and $\theta_{0k}$ be the same quantities for the two historical (adult) groups respectively.
\citet{o2001bayesian} introduced the notion of using two different prior distributions in clinical trial settings:
a {\it design prior}, a more realistic choice used to evaluate the likely properties of a design, and
an {\it analysis prior}, a typically more conservative choice that will actually be used when the data are observed.
Consider the latter choice first;
our analysis prior uses a
{\it commensurate prior} framework \citep{hobbs2011hierarchical}, and assumes
\beq\label{commensprior}
\theta_k | \theta_{0k} \sim N(\theta_{0k}, 1/\tau_k),
\mbox{ and } \theta_{0k} \sim N(0, \sigma_0^2), \; k=1, 2 ,
\eeq
where we assume $\theta_{0k}$ follows a vague prior for the adult percent spleen reductions, e.g. $\sigma_0 = 100$ is known.
The commensurability parameters (precisions) $\tau_k$ are assigned independent hyperpriors, e.g., the conjugate choices $\tau_k \sim G(1/50, 1)$ as 
relatively vague Gamma specifications.

Turning to the observed data,
suppose we assume that $Y_{kj}$ and $Y_{0kj}$, the observed percent reductions for each pediatric and adult patient,
are also normally distributed, that is,
\beq\label{data}
Y_{kj} \stackrel{ind}{\sim} N(\theta_{k}, 1/\omega) \mbox{ and }
Y_{0kj} \stackrel{ind}{\sim} N(\theta_{0k}, 1/\omega_0) \; ,
\eeq
where the patient index $j$ runs from 1 to $n_k$ or $n_{0k}$, respectively,
and we again use vague conjugate priors for $\omega$ and $\omega_0$; e.g.,
$\omega, \omega_0 \stackrel{iid}{\sim} G(1/100 , 1)$.

We design and calibrate the trial to have acceptable long-run frequentist properties (Type I error and
power) for {\it any} given value of $n'$, and then select the value that maximizes trial payoff (as defined below).
Also, our design uses
one interim look to check for early stopping due to success or futility,
accounting for commensurability of the adult and pediatric data,
but does {\it not} consider adjusting
the randomization ratio depending on how many adults we are ``effectively" borrowing.
Such an enhancement would be possible to add to our design \citep{hobbs2013adaptive,normington2018}.
Finally, we note that either posterior or predictive
distributions can be used for these calculations.
For simplicity, we use the former (implemented via MCMC computation 
in {\tt BUGS, R/Stan} or {\tt SAS}) for our interim stopping rules as follows:

\bdes
\item[Early winner:]
If at the interim look, the probability that the novel treatment arm ($k=2$) is better exceeds some
prespecified probability $p_U$, i.e., if
$$
P(\theta_2 > \theta_1 | Data) > p_U \; ,
$$
then Arm 2 is declared the {\it early winner} and the trial is stopped early.
We might take $p_U$ be a fairly high value, so early trial termination is permitted only when evidence for an early winner is overwhelming \cite{anderson2014timing}.

\item[Final winner:]
Early winner rules are typically paired with corresponding {\it final winner} rules, e.g:
If, after all patients have been randomized and reported results, the probability that the treatment arm is the best exceeds some
prespecified probability $p_0$, i.e., if
$$
P(\theta_2 > \theta_1 | Data) > p_0 \; ,
$$
then Arm 2 is declared the {\it final winner}.  If however the treatment arm cannot meet
this criterion, then we do not make a final selection as to ``best treatment", and merely
summarize the performance of both treatments.
We might set $p_0$ as a slightly less demanding threshold than the early winner level $p_U$.

\item[Early futility:]
If at the interim look, the probability that the novel treatment arm ($k=2$) is better than some prespecified minimally
tolerable response rate $\theta_{min}$ falls below some
prespecified probability $p_L$, i.e., if
$$
P(\theta_2 > \theta_{min} | Data) < p_L \; ,
$$
then the trial is declared {\it futile} and is stopped early (i.e., after just $n'$ patients).
We might set $\theta_{min}$ as the minimum reduction in spleen volume from the novel treatment that can be clinically relevant),
and take $p_L$ fairly small.  
Thus, if the treatment cannot muster at least a $p_L$ chance of a  $\theta_{min}\%$ reduction in spleen volume at our interim look, we
will give up on the treatment and the trial is stopped early for futility.
\edes

\noindent
\textbf{Algorithm:}
In summary, our overall algorithm for given choices of $n$ and $n'$ is as follows:

\begin{enumerate}

\item
Fix $\theta_1 = \theta_2 = 0$,
so that the null hypothesis is true (no difference in pediatric spleen volume reduction between treatment and placebo).

\item
Use equation (\ref{data}) with a fix $w$ to
generate Monte Carlo pediatric observations $Y_{kj}, \: j=1, \ldots, n'_k, \: k=1,2$,
and combine with the actual adult observations $Y_{0kj}, \: j=1, \ldots, n_{0k}, \: k=1,2$.

\item Perform the interim look at the data, estimating the posterior precision of the pediatric response in each group using both the pediatric data alone and the full model (commensurate prior with adult historical data);
    namely, ${Prec}(\theta_k \vert \bD')$ and ${Prec}(\theta_k \vert \bD', \bD_0)$, where $\bD'$ and $\bD_0$ denote the interim pediatric and full adult data, respectively.
    If posteriors are being computing using MCMC, these precisions would just be the reciprocals of the sample variances of the $G$ MCMC samples $\{\theta_k^{(g)}, g=1, \ldots, G\}$ for the 2 groups and the 2 different models (interim pediatric only vs.\ full data).

\item For k=1,2, compute the effective historical sample sizes
  $$
  {EHSS}_k = min \left ( max\left[n_{0k} \left(\frac{Prec(\theta_k \vert \bD', \bD_0)}{Prec(\theta_k \vert \bD')} - 1 \right ), 0 \right ], n_{0k} \right )\; ,
  $$
so that $EHSS_1 + EHSS_2$ is the total effective historical sample size \citep{hobbs2013adaptive}. We recommend monitoring the $EHSS$ 
to ensure it is not unacceptably large; say, more than twice as large as $n'$, the interim pediatric sample size.
Such a check reflects a guideline that a regulatory authority such as the FDA might impose, requiring 
that a significant proportion of the total information used in a pivotal trial comes from the trial data itself, not from historical information.  Note that such restrictions could be imposed on $EHSS_1$ and $EHSS_2$ separately; e.g., requiring more caution when borrowing from historical cases as opposed to historical controls \citep{li2019target}.

\item
Use the early winner and futility rules above to see if the trial can stop now; if so, write this down and skip the next step.


\item
Use (\ref{data}) with the same $w$ in Step 2 to
generate the remaining pediatric observations $Y_{kj}, \: j=n'_k+1, \ldots, n_k, \: k=1,2$,
and then use the ``final winner" rule above to
  see if the trial can now choose a definite winner.  Note that this approach is equivalent to using an
  appropriately sized Bayesian credible interval (BCI) for the pediatric treatment effect $\Delta \equiv \theta_2 - \theta_1$.
   For example, with $p_0 = 0.975$, the equivalence would be a 95\% equal-tail BCI: if it is totally above 0, conclude treatment is superior to placebo; if it is totally below 0, conclude treatment is inferior to placebo; and if it contains 0, fail to conclude superiority of either treatment. The equivalence of $p_U$ and $p_L$ to their corresponding BCIs can be established respectively.

\item
Repeat Steps 2--6 $N_{rep}$ times, and estimate the Type I error of our design as
\beq\label{power}
 \frac{\# \mbox{ of treatment early winners} + \# \mbox{ of treatment final winners}}{N_{rep}} \; .
\eeq
Repeat Steps 2--6 and grid search on the choices of $p_L, p_U$, and $p_0$ in the stopping rules for the study designs with the desired test size (say, $5.0\%$).

\item
Keep $\theta_1 = 0$ but change $\theta_2 = 20$ (or any known value meet the target efficacy),
so that now the alternative hypothesis is true (clinically significant improvement in pediatric spleen volume reduction
on treatment as compared to placebo).
Repeat Steps 2--7 above, estimating the power of our design
using equation (\ref{power}), and check if it is above the desired level (e.g., $80.0\%$).
If the power is not above the desired level, we can alter the choices of $p_L, p_U$, and $p_0$ in the stopping rules
and try again, however, to maintain the procedure's Type I error calibration we can only choose $p_L, p_U$, and $p_0$ among the study designs with the desired test size obtained in Step 6.
(Otherwise, if no design obtained in Step 6 achieves the desired power, we might instead need to increase $n$, or alter the hyperpriors on the $\tau_k$ so that more strength is borrowed from the historical adult data).

\item
Rather than fix $\theta_1$ and $\theta_2$ as in Steps 1 and 8,
repeatedly {\it sample} them from a particular {\it design prior}, for example
\beq\label{designprior}
\theta_1 \sim N(\theta_{1,des}, \sigma_{1,des}^{2}) \mbox{ and } \theta_2 \sim N(\theta_{2,des}, \sigma_{2,des}^{2}) 
\eeq
for the children where $\sigma_{1,des}$ and $\sigma_{2,des}$ are known, and set $\theta_{1,des}  = 0$ and $\theta_{2,des}  = \Delta$. 
We again use the actual adult observations, and to be realistic we might set $\Delta$ smaller than the mean observed reduction in adults, to reflect the plausible situation that the treatment offers a greater benefit to adults than it does to children.  

In all cases, we repeat Steps 2--7 above again, estimating the marginal probabilities of early stopping $\hat{P}$ under our design prior, including early futility and early winner under the design prior. 
The numerator (benefit) of the payoff function can be defined under both the null hypothesis and alternative hypothesis for optimizing beneficial goals (i.e., to estimate the marginal probabilities of making correct decisions at IA, including early futility under null hypothesis and early efficacy under the alternative hypothesis separately).  Define
\beq\label{P1P2Xu}
\hat{P}_1 = \frac{\# \mbox{ of early futility stops}}{N_{rep}} \mbox{ under $H_0$ }
\mbox{ and }
\hat{P}_2 = \frac{\# \mbox{ of treatment early winners}}{N_{rep}} \mbox{ under $H_a$} \; .
\eeq

Use these quantities to compute the trial payoff as
\beq
\label{Payoff}
\mbox{Payoff} = \frac{w \hat{P}_1 + (1-w) \hat{P}_2}{\hat{P} n' + (1-\hat{P}) n} \; ,
\eeq
where $w \in (0,1)$ is a preselected weight that trades off the two types of decisions in (\ref{P1P2Xu}).
The denominator (cost) can be explained as the expected sample size of the study design. 

An alternative fully Bayesian payoff function computes the marginal probabilities of early stopping, early futility and early efficacy under the design prior.  This redefines
$\hat{P}_1$ and $\hat{P}_2$ as 
\beq\label{P1P2Bayes}
\hat{P}_1 = \frac{\# \mbox{ of early futility stops}}{N_{rep}}
\mbox{ and }
\hat{P}_2 = \frac{\# \mbox{ of treatment early winners}}{N_{rep}} \mbox{ both under the design prior}\;,
\eeq
now averaging over the design prior, and again use these quantities to compute the trial payoff in (\ref{Payoff}).

\item
Repeat {\it all} the steps above (Steps 1--9) across a  grid of $n'$ values.
Choose the $n'$ value that maximizes the Payoff as computed in equation (\ref{Payoff}).
This $n'$ is optimal under design prior (\ref{designprior}), and the resulting design has correctly calibrated and acceptable Type I error and power by construction.
\end{enumerate}

We have created an {\tt R} package,
{\bf optimIA}, available at \url{https://github.com/wxwx1993/Bayesian_IA_Timing}, to implement our algorithm. In the next section, we use the algorithm to determine the optimal timing for an interim analysis in the context of our pediatric study setting.


\section{Simulation study}\label{results}

\subsection{Simulation settings}
In our simulation study, we implement the Bayesian algorithm proposed in Section~\ref{method}. To illustrate the method, we simulated the historical study data hypothetically from a normal distribution with the endpoint 
being \% reduction in spleen volume. The historical study sample size are assumed to be $n_{01} + n_{02} = 50$, in which $n_{01}=25$ are in the placebo group and $n_{02}=25$ are in the treated group. The simulated historical data have mean difference $\Delta_0 = 25$ and corresponding
standard deviation $\text{SD} = 22$. 
We consider a current pediatric  study with planned sample size of 40 ($n_1=n_2=20$ per arm). This sample size will provide approximately $81.2\%$ power (under one-sided $5.0\%$ Type I error)  to detect a target treatment difference of $H_a: \theta_2 - \theta_1 = 20$ without considering interim look or external evidence (i.e. historical data borrowing in this study). We 
consider potential IA times after $n' = 0, 4, 8, 12, 16, 20, 24, 28, 32, 36$ and 40 enrollments. 
We vary the weight as $w = 0, 0.5, 0.75, 1$, to represent the considerations mentioned in Section~\ref{method}. 
For example, with $w=0$, the benefit (numerator of the payoff function) will represent the probability of an early win when the true treatment effect is at the target; stopping early for futility is deemed to have no value.  
In this case, the highest payoff will maximize the probability of early success at IA when the drug is working (benefit), while controlling the expected sample size (cost).  The use of $w=0.5$ will place equal weight on stopping early for a win and stopping early to give up on the drug,
while $w=0.75$ places heavier emphasis on earlier abandonment of
an apparently ineffective drug. This may occurs if the external information suggested less favorable profile of the drug or emergence of a new competitor drug make the new treatment less desirable for further development. The use of $w=1$ is extremely unlikely in practice as it will place no benefit on an early win, and we present the outcome only for completeness.

We set the null hypothesis and target alternative hypothesis as, respectively, 
\begin{align*}
    &H_0:  \theta_2 - \theta_1 =0 \\
  \mbox{and} \ \  &H_a:  \theta_2 - \theta_1 =20 \; .
\end{align*}
We choose a minimal efficacy level of $\theta_{min} = 15$ for 
defining futility at the interim look.
Under our design prior, we consider four values of the mean treatment effect:  $\Delta = 0$, for scenarios in which we expect the new treatment will show no improved efficacy; $\Delta = 15$, for scenarios 
in which we expect the treatment will achieve minimal efficacy; 
$\Delta = 25$, for scenarios in which we expect the treatment
will achieve the same high efficacy as the adult (historical) study; 
and finally $\Delta = 35$, for surprising scenarios in which we expect the new treatment will achieve even higher efficacy than that seen in the historical adult study. The number of replicates for calculating and calibrating the Type I errors and powers within each simulation scenario is 5000. For each MCMC run, we specify a chain of 5000 iterations, with the first $20\%$ of the samples deleted as the initial ``burn-in" period.

\subsection{Simulation results}

\begin{figure}[H]
\centering
\includegraphics[height=0.35\textheight]{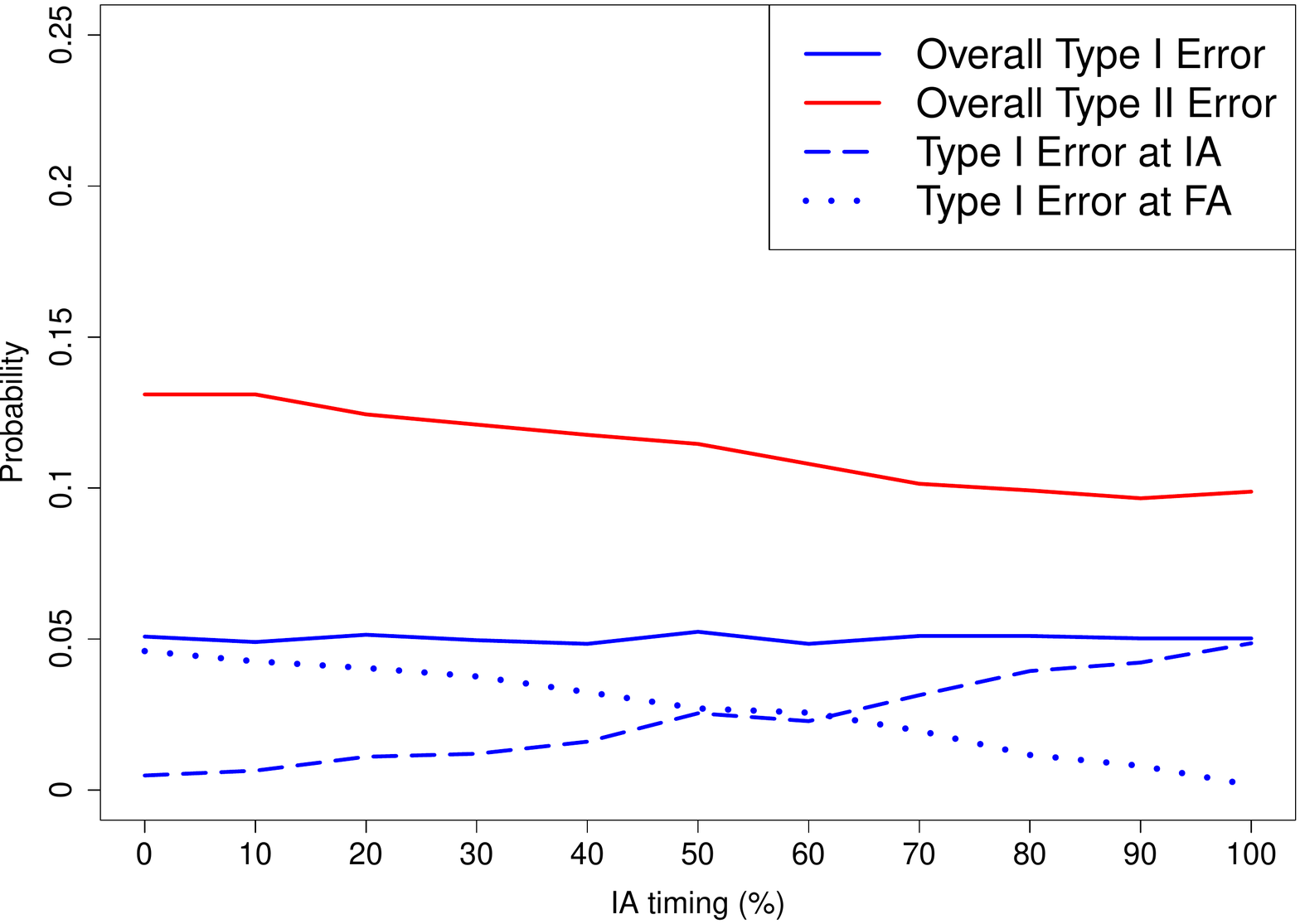}
\caption{IA timing (\%) vs Type I and Type II errors. We calibrate Type I error (one-sided at the size of $5.0\%$) by grid searching the suitable $p_u$ value and fix $p_0 = 0.975$ and $p_l = 0.25$.  By choosing suitable stopping rules at the design stage, the Type II errors are also controlled at the level of $13.1\%$ ($9.7\%-13.1\%$), which is equivalent to $86.9\%-90.3\%$ power. The blue dashed/dotted lines show the partition of overall Type I error into that due to treatment early winners at interim analysis (IA) and treatment final winners at final analysis (FA) throughout different IA times.  
}
\label{TypeI}
\end{figure}

Figure~\ref{TypeI} shows our algorithm controls power at the level of $86.9\% \ (86.9\%-90.3\%)$ when we calibrate the overall Type I error (one-sided at $5\%$) by finding the suitable choice of $p_U$
where we fix $p_0 = 0.975$ and $p_L = 0.25$. 
This represents a notable boost of power (at least $5.7\%$) compared to the standard frequentist method without historical data borrowing. Figure~\ref{TypeI} also shows the partition of the overall Type I error into that 
contributed by treatment early winners at IA and that arising from final winners at FA (blue dashed and dotted lines, respectively).  We see the component due to early winners is elevated for later IA times, yet the overall Type I error is always controlled by construction. Table~\ref{borrow_info} shows the amount of EHSS borrowed from the historical study under different design priors. Noting that in general we borrow more placebo than the treated, since under the design prior, we tend to believe the placebo arms are similar between adults and pediatrics, since they are both untreated and should have no reduction on spleen volume. Yet the amount of borrowing for treated arm highly depends on the specifications of design priors and true treatment effect; note for example the extensive borrowing from treateds even for later IA times when $\Delta=25$ in Table~\ref{borrow_info}. 
If on the other hand we believe the effects of drug on children are quite different from those in adults (other values of $\Delta$ in the table), we tend to rely less on the historical data due to incommensurability. This is the reason that the borrowing from the treatment arm is maximized at $\Delta=25$, which is the fully commensurate case here; for $\Delta=35$, treatment arm borrowing drops again to levels comparable to that seen for $\Delta=15$. In addition, we observe the amount of borrowing decreases as the IA time goes up. This is likely because later IA times cast greater doubt on commensurability. We also find the historical sample sizes (HSS) impact EHSS. Briefly, the borrowing proportions (EHSS/total HSS) tend to be higher for larger HSS. Additional simulation results under different sizes of historical adult data sets are presented in Appendix~\ref{EHSS}. 

Figure~\ref{payoff} and Table~\ref{payoff_tab1} give our main results for payoff function defined in (\ref{P1P2Xu})-(\ref{Payoff}). Figure~\ref{payoff} presents the values of payoff functions under different scenarios. It is clear that each estimated payoff curve has a maximal point, which can be interpreted as the optimal time for an interim analysis. In general, we observe the optimal IA times using the specified payoff function are within the range of recruiting $50\%-80\%$ of patients, depending on the choices of design priors. One interesting perspective is that when $\Delta = 15$, in which we expect the treatment only achieves minimal efficacy, the study design provides the latest IA time compared to other scenarios. This is not surprising as under this marginal case where the current study achieves only minimal efficacy, more data (longer waiting time) is needed in order to make a clear decision 
regarding early wins and losses.  
That is, this is the case that is most difficult for our 
commensurate prior framework to handle, as the decision whether or not to borrow is not clear-cut.  
In contrast, under the surprised high efficacy scenario ($\Delta=35$), we observe the earliest optimal IA timing, reflecting the investigator's high confidence in the treatment's effectiveness in the current pediatric trial. 
The no efficacy ($\Delta = 0$) and high efficacy ($\Delta = 25$) scenarios provide roughly identical optimal IA timing, yet it's worth noting that the stopping mechanisms of IA are different: under no efficacy, early futility dominates the IA decision, while under high efficacy, the early stops are due to early winners. 

Table~\ref{payoff_tab1} also illustrates the impact of the weight $w$ on our pediatric study design. 
We see a  trend towards earlier optimal times when $w$ is larger for all scenarios, which corresponds to our placing greater importance on early stopping for futility. In practice, the choice of $w$ is somewhat subjective and thus requires external information, probably from elicited expert opinions \citep{hampson2014bayesian}.  Extra caution needs to be paid in certain scenarios; e.g., when the rates of change over IA time in the probabilities of early win and early futility are very different, the choice of $w$ may have larger impact on the optimal IA timing.  Results using the alternative, fully Bayesian payoff function (\ref{P1P2Bayes})
are presented in Appendix~\ref{fullbayes}.

Figure~\ref{cost} plots the expected sample size for the implementations of our study under different design priors. Expected sample size was defined as the expectation of sample size as we either conduct the IA and then stop, or conduct the IA and then continue to recruit more patients until the completion of study. The finding is that the minimal expected sample size appears if we conduct the interim analysis when have recruited $40\%-50\%$ patients. 
Our findings, under a Bayesian adaptive design, shows the timing of minimizing expected sample size appears earlier than those shown in \citet{TogoIwasaki2013}, likely due to the information borrowed from the historical data. However, as demonstrated in Section~\ref{method}, minimizing the expected sample size is not necessary to be the only criterion for finding optimal IA timing.
Rather, a view toward maximizing a payoff function that characterizes the benefit/cost ratio is used in our Bayesian adaptive design. Under the optimal IA timing obtained by maximizing our specified payoff function, $13.8$ to $44.9\%$ savings in expected sample size were observed. This often outperforms the study designs presented in \citep{TogoIwasaki2013}, which found $18$ to $22\%$ savings in sample size with a single IA, even if the primary goal of our design is not to maximize savings in expected sample size.
Again, the reason is our Bayesian adaptive design effectively borrows from historical information.
We also note that our simulations reveal enormously larger
Monte Carlo standard errors (SEs) associated with the 
expected sample size estimation for early IA times
(left side of Figure~\ref{cost}),
yet the SEs shrink to 0 as IA time increases, reflecting the fact that later IAs provide progressively more accurate estimation of cost as they progress toward full enrollment. The data tables to construct Figures~\ref{TypeI}-\ref{cost} are provided in Appendix~\ref{datatable}. 

\section{Discussion and Future Work}\label{discussion}

In this paper, we have used a Bayesian commensurate prior formulation to design a clinical trial with an optimally placed single interim look.  Our goal was to move beyond simple optimality criteria that involve only overall expected sample size to those that actually measure not just the savings resulting from persons not enrolled in the study, but gains to the sponsor arising from making a correct decision as soon as possible.  While the goal of an optimally designed clinical trial is to be able to declare study success as soon as possible when the drug is working, it's also important to stop the trial as soon as possible when the drug is not working. The use of weight in the payoff function allows the project team to assess the relative importance of these two actions and reach to an ethical decision based on existing information.  Also, since the design has controlled both Type I and Type II errors overall, the proposed payoff function was intended to maximize the chance of making a ``weighted" correct decision as high as possible and as early as possible.  

Our findings suggest optimal IA times tended to be different from the optimal time based on cost alone (when the expected sample size is the smallest).  The optimal time may be earlier when the treatment effect is unequivocal (either very small or very large), or when greater importance is placed on early stopping for futility (higher $w$ values). 
By contrast, equivocal treatment effects (i.e., close to those deemed minimally clinically significant) or a higher emphasis on early stopping for 
efficacy (the ``early winner") lead to later optimal IA times.

Our payoff function (\ref{Payoff}) resulted from a hybrid of Bayesian and frequentist ideas, which we do not view as inappropriate in a field where methods that are formally Bayesian but also required to have good frequentist properties are routinely used. Unlike the existing methods which calculate cost only based on a fixed alternative hypothesis, the current estimate of cost (denominator) based on different Bayesian prior allows a more realistic estimation of the cost. For an actual trial design, the expected cost should be assessed under all possible scenarios (based on existing knowledge) to assess their impact on optimal IA timing. 

Still, while useful, our payoff function is fairly ad hoc, namely
through equation (\ref{Payoff})
where
$\hat{P}_1 = (\# \mbox{ of early futility stops})/{N_{rep}}$ is an estimate of the probability
of an ``early loss",
$\hat{P}_2 = (\# \mbox{ of treatment early winners})/{N_{rep}}$ is a corresponding estimate of the
probability of an ``early win",
$\hat{P}$ is the probability of early stopping for any reason,
and $w \in (0,1)$ is a weight that trades off these two early stop probabilities.
 
Suppose we define two more Bayesian posterior probability estimates \cite{cheng2007optimal},
$$
\hat{P}_3 = \frac{\# \mbox{ of treatment late winners}}{N_{rep}}
\mbox{ and }
\hat{P}_4 = 1 - \sum_{k=1}^3 \hat{P}_k
= \frac{\# \mbox{ of treatment late losers}}{N_{rep}} \; .
$$
Let us now think of gain and cost on a purely financial (i.e., dollar) scale.  Obviously we would need help from
the trial sponsor to do this, but we could consider a range of possibilities.
Define
\begin{align*}
    &\mbox{Gain(early loss)} = a_1 , \;
\mbox{Gain(late loss)} = a_2 < a_1,\\
  \mbox{and }  &\mbox{ Gain(early win)} = b_1, \;
\mbox{Gain(late win)} = b_2 < b_1 \; .
\end{align*}
We might take $a_1=0$, so that the ``gain" from a late loss $a_2$ is actually negative, corresponding
to the financial loss associated with having to postpone development of
other drugs while we waited for this one to fail.
Similarly, we would surely take $b_1 > 0$, but would also take $0 < b_2 < b_1$, due to the missed
opportunity to sell the drug while we waited for the trial to run to completion.

Next, let $C$ be the trial's per-patient cost.  Then we would have
$
\mbox{Cost(early loss)} = \mbox{Cost(early win)} = Cn',
$ and 
$
\mbox{Cost(late loss)} = \mbox{Cost(late win)} = Cn  ,
$
since patients cost the same regardless of whether we win or lose.
Thus the Bayesian expected net gain for the trial is:
$$
\mbox{E(Gain -- Cost)} =
\hat{P}_1 a_1 + \hat{P}_2 b_1 + \hat{P}_3 b_2 + \hat{P}_4 a_2
- C [(\hat{P}_1 + \hat{P}_2)n' + (\hat{P}_3 + \hat{P}_4)n] \; .
$$
Once again, we could choose the location of the IA to maximize this posterior expected net gain,
instead of the Payoff function in (\ref{Payoff}).  
We hope such an investigation will be the subject of a future manuscript.

\section*{acknowledgements}
The work of all three authors was supported in part by Sanofi Pharmaceuticals.  
The authors are grateful to Dr.\ Xun Chen for initial discussions that influenced the direction of this work. The computations in this paper were run on the Odyssey cluster supported by the FAS Division of Science, Research Computing Group at Harvard University. We thank the referees for their careful reading of our work and for their thoughtful and helpful comments.

\section*{conflict of interest}
None



\bibliographystyle{ama}
\bibliography{ama}

\newpage

\section*{Figures and tables}

\begin{table}[H]
\centering
\caption{ Effective historical sample sizes (EHSS) borrowed from historical study at interim look (Placebo/Treated) when the total historical sample size (HSS) is 50 (25/25)}
\label{borrow_info}
\begin{tabular}{ccccccc}
IA time &{  $\Delta = 0$ } & {  $\Delta = 15$} & {  $\Delta = 25$} & { $\Delta = 35$}    \\ \hline
  30\% &   19.86 / 10.11& 19.94 / 17.75 & 20.09 / 20.17 & 20.01 / 17.62  \\ 
  40\% &   19.61 / 7.55 & 19.74 / 16.64 & 19.72 / 19.88 & 19.63 / 16.26  \\ 
  50\% &   19.25 / 5.73 & 19.35 / 15.30 & 19.42 / 19.49 & 19.31 / 15.11  \\ 
  60\% &   18.59 / 4.63 & 18.81 / 14.15 & 18.81 / 18.81 & 18.81 / 13.83  \\ 
  70\% &   18.18 / 3.91 & 18.29 / 12.85 & 18.22 / 18.32 & 18.14 / 12.65  \\ 
  80\% &   17.42 / 3.47 & 17.44 / 11.74 & 17.67 / 17.65 & 17.48 / 11.63  \\ 
  90\% &   16.73 / 3.08 & 16.68 / 10.66 & 16.96 / 16.94 & 16.76 / 10.79  \\ 
   \hline
\end{tabular} 
\end{table}

\begin{figure}[H]
\centering
\includegraphics[height=0.6\textwidth]{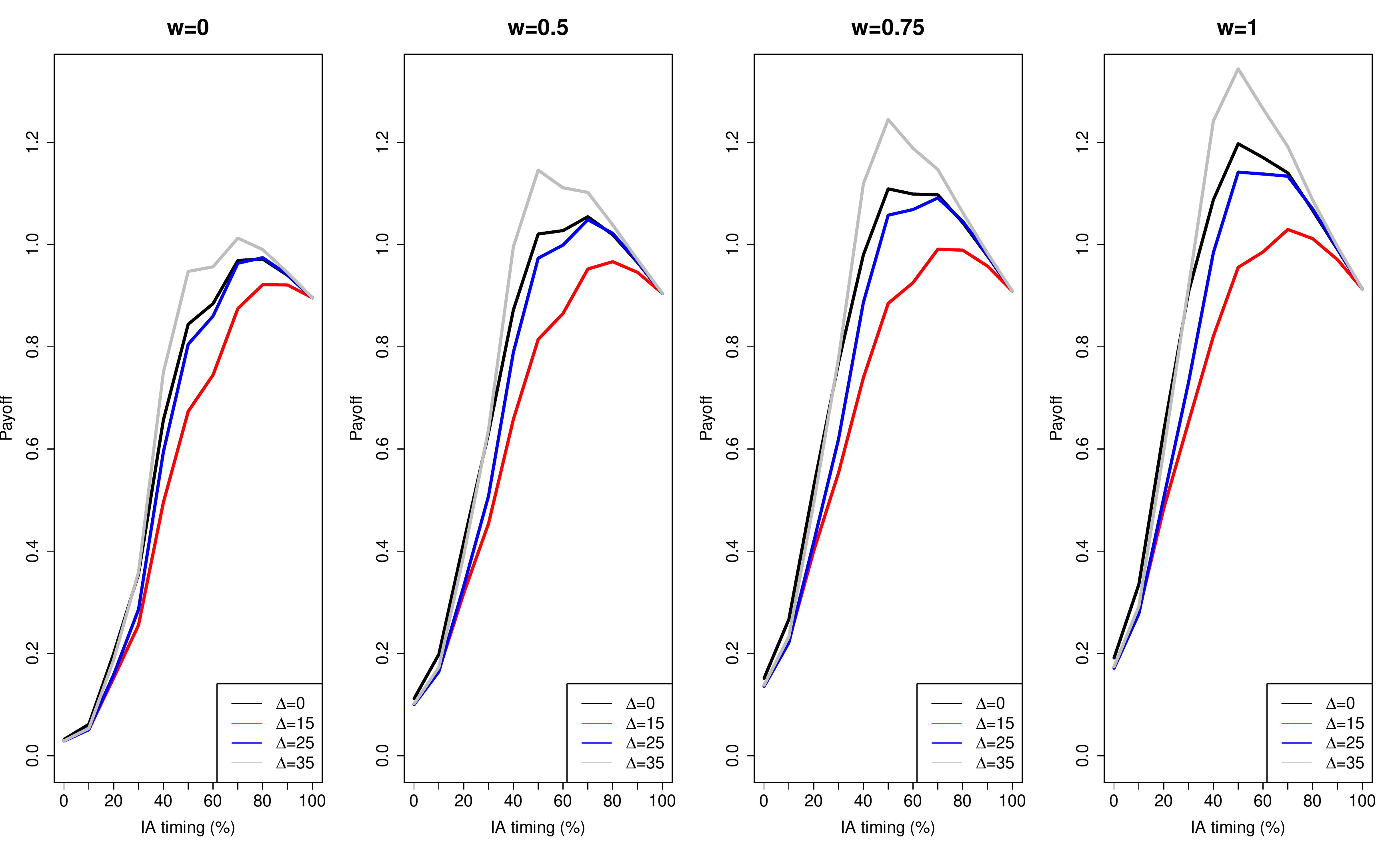}
\caption{IA timing $(\%)$ vs Payoff. Each panel represents the values of payoff function with respect to different weights $w=0,0.5,0.75,1$. IA timing under different design prior with effect size: 1) $\Delta=0$, no efficacy, 2) $\Delta=15$, at minimal efficacy, 3)  $\Delta=25$, high efficacy, and 4) $\Delta=35$, surprised high efficacy. 
}
\label{payoff}
\end{figure}

\begin{table}[H]
\centering
\caption{\label{sim:results} Optimal IA timing chosen under different choices of design priors and choices of weights in the payoff}
\label{payoff_tab1}
\begin{tabular}{cccccc}
effect size & $w=0$ & $w=0.5$ & $w=0.75$ & $w=1$ \\ \hline
0    & 32 (80\%)  & 28 (70\%) & 20 (50\%)  & 20 (50\%)    \\  \hline
15   & 32 (80\%)  & 32 (80\%) & 28 (70\%)  & 28 (70\%)  \\ \hline
25   & 32 (80\%)  & 28 (70\%) & 28 (70\%)  & 20 (50\%) \\ \hline
35   & 28 (70\%) & 20 (50\%)  & 20 (50\%)  & 20 (50\%)    \\ 
   \hline 
\end{tabular} 
\end{table}

\begin{figure}[H]
\centering
\includegraphics[height=0.35\textheight]{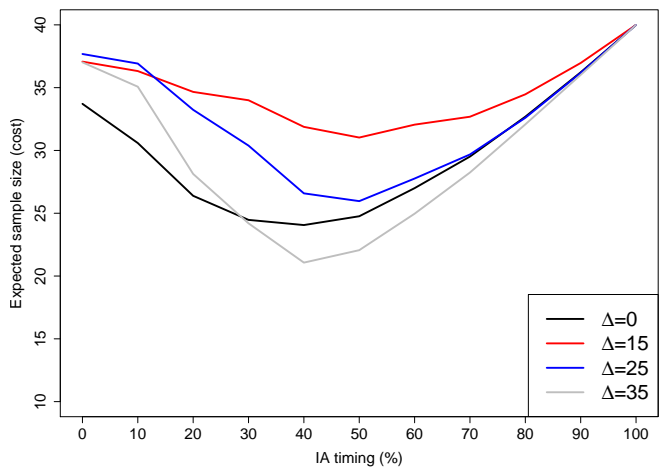}
\caption{IA timing $(\%)$ vs Expected sample size. Expected sample size was defined as the expectation of sample size as we either conduct the IA and then stops or conduct the IA and then continue to recruit patients until the completion of the study. Under the optimal IA timing obtained by maximizing our specified payoff, $13.8\%-44.9\%$ saves of expected sample size were observed.
}
\label{cost}
\end{figure}

\newpage
\section*{Appendix}

\setcounter{table}{0}
\setcounter{figure}{0}
\setcounter{subsection}{0}
\renewcommand\thetable{A\arabic{table}}
\renewcommand\thefigure{A\arabic{figure}}
\renewcommand\thesubsection{A\arabic{subsection}}

\subsection{Additional simulations for effective historical sample sizes (EHSS)}
\label{EHSS}
 We change the total sample size ($n_{01} + n_{02}$) of historical adult data sets from 50 to 20 and 100, to see the impact on EHSS.  We also experimented with keeping this total sample size at 50, but switching from an equal allocation between treated and controls ($n_{01} = n_{02} = 25$) to unequal allocations, namely $n_{01} = 10, n_{02} = 40$ and 
$n_{01} = 40, n_{02} = 10$.  The results indicate that, in general, we borrow more from the placebo than the treated group in all scenarios.  This is because under the design prior, we tend to believe the placebo arms are similar between adults and pediatrics, since they are both untreated and should thus have no reduction in spleen volume.  However, the amount of borrowing for the treated arm depends highly on the specifications of the design priors and the true treatment effect. The absolute amount of borrowing (as measured by EHSS) is also sensitive to the total historical sample sizes (HSS), and the borrowing proportions (EHSS/total HSS) also tend to be higher for larger HSS. In addition, the trend is that the amount of borrowing decreases as the IA time goes up, which is sensible since later IA times cast greater doubt on commensurability.
\begin{table}[H]
\centering
\caption{ Effective historical sample sizes (EHSS) borrowed from historical study at interim look (Placebo/Treated)  when the (balanced) total historical sample size is 20 (10/10)}
\label{borrow_infoa1}
\begin{tabular}{ccccccc}
IA time &{  $\Delta = 0$ } & {  $\Delta = 15$} & {  $\Delta = 25$} & { $\Delta = 35$}    \\ \hline
  30\% & 7.18 / 3.46 & 7.31 / 6.59 & 7.42 / 7.45 & 7.35 / 6.54 \\ 
  40\% & 6.57 / 2.40 & 6.80 / 5.86 & 6.83 / 6.91 & 6.74 / 5.72 \\ 
  50\% & 5.84 / 1.66 & 6.03 / 4.99 & 6.17 / 6.18 & 6.06 / 4.95 \\ 
  60\% & 5.06 / 1.24 & 5.34 / 4.31 & 5.41 / 5.41 & 5.34 / 4.22 \\ 
  70\% & 4.51 / 0.94 & 4.76 / 3.71 & 4.79 / 4.84 & 4.73 / 3.63 \\ 
  80\% & 4.02 / 0.79 & 4.23 / 3.29 & 4.32 / 4.34 & 4.23 / 3.21 \\ 
  90\% & 3.68 / 0.64 & 3.84 / 2.91 & 3.97 / 3.93 & 3.83 / 2.89 \\ 
   \hline
\end{tabular} 
\end{table}

\begin{table}[H]
\centering
\caption{ Effective historical sample sizes (EHSS) borrowed from historical study at interim look (Placebo/Treated)  when the (balanced) total historical sample size is 100 (50/50)}
\label{borrow_infoa2}
\begin{tabular}{ccccccc}
IA time &{  $\Delta = 0$ } & {  $\Delta = 15$} & {  $\Delta = 25$} & { $\Delta = 35$} \\   \hline
  30\% & 40.76 / 22.60 & 40.86 / 36.39 & 41.09 / 41.30 & 40.97 / 36.09 \\ 
  40\% & 40.65 / 18.41 & 40.72 / 34.41 & 40.69 / 41.06 & 40.53 / 33.70 \\ 
  50\% & 40.53 / 15.33 & 40.62 / 32.14 & 40.67 / 40.81 & 40.54 / 31.70 \\ 
  60\% & 40.03 / 13.48 & 40.19 / 30.14 & 40.14 / 40.15 & 40.22 / 29.54 \\ 
  70\% & 40.07 / 12.36 & 40.08 / 27.82 & 39.85 / 40.04 & 39.88 / 27.65 \\ 
  80\% & 39.66 / 11.65 & 39.59 / 26.15 & 39.67 / 39.67 & 39.04 / 24.52 \\ 
  90\% & 39.17 / 10.93 & 39.14 / 24.24 & 39.28 / 39.20 & 38.41 / 22.84 \\ 
   \hline
\end{tabular} 
\end{table}

\begin{table}[H]
\centering
\caption{ Effective historical sample sizes (EHSS) borrowed from historical study at interim look (Placebo/Treated)  when the (unbalanced) total historical sample size is 50 (10/40)}
\label{borrow_infoa3}
\begin{tabular}{ccccccc}
IA time &{  $\Delta = 0$ } & {  $\Delta = 15$} & {  $\Delta = 25$} & { $\Delta = 35$}    \\   \hline
  30\% & 7.50 / 15.20 &7.55 / 28.09 & 7.62 / 32.32 & 7.57 / 27.87 \\ 
  40\% & 7.17 / 10.95 &7.27 / 26.19 & 7.24 / 31.97 & 7.21 / 25.62 \\ 
  50\% & 6.75 / 8.18 & 6.79 / 24.04 & 6.84 / 31.46 & 6.80 / 23.70 \\ 
  60\% & 6.22 / 6.58 & 6.30 / 22.17 & 6.34 / 30.60 & 6.30 / 21.66 \\ 
  70\% & 5.78 / 5.55 & 5.85 / 20.08 & 5.83 / 30.21 & 5.82 / 19.88 \\ 
  80\% & 5.30 / 4.99 & 5.34 / 18.42 & 5.40 / 29.54 & 5.34 / 18.24 \\ 
  90\% & 4.90 / 4.44 & 4.91 / 16.69 & 5.04 / 28.71 & 4.91 / 16.95 \\ 
   \hline
\end{tabular} 
\end{table}

\begin{table}[H]
\centering
\caption{ Effective historical sample sizes (EHSS) borrowed from historical study at interim look (Placebo/Treated)  when the (unbalanced) total historical sample size is 50 (40/10)}
\label{borrow_infoa4}
\begin{tabular}{ccccccc}
IA time &{  $\Delta = 0$ } & {  $\Delta = 15$} & {  $\Delta = 25$} & { $\Delta = 35$}    \\    \hline
  30\% & 34.20 / 5.91 & 34.28 / 7.93 & 34.49 / 8.57 & 34.42 / 7.93 \\ 
  40\% & 34.12 / 5.10 & 34.25 / 7.68 & 34.30 / 8.49 & 34.14 / 7.55 \\ 
  50\% & 34.08 / 4.36 & 34.23 / 7.29 & 34.30 / 8.33 & 34.18 / 7.23 \\ 
  60\% & 33.71 / 3.83 & 33.70 / 6.43 & 33.89 / 8.02 & 33.88 / 6.79 \\ 
  70\% & 33.70 / 3.35 & 33.34 / 5.98 & 33.63 / 7.72 & 33.57 / 6.34 \\ 
  80\% & 33.22 / 3.03 & 32.83 / 5.52 & 33.42 / 7.32 & 33.23 / 5.88 \\ 
  90\% & 32.71 / 2.72 & 32.12 / 5.14 & 32.26 / 6.45 & 32.78 / 5.52 \\ 
   \hline
\end{tabular} 
\end{table}

\subsection{Fully Bayesian payoff function}
\label{fullbayes}
\begin{figure}[H]
\centering
\includegraphics[height=0.6\textwidth]{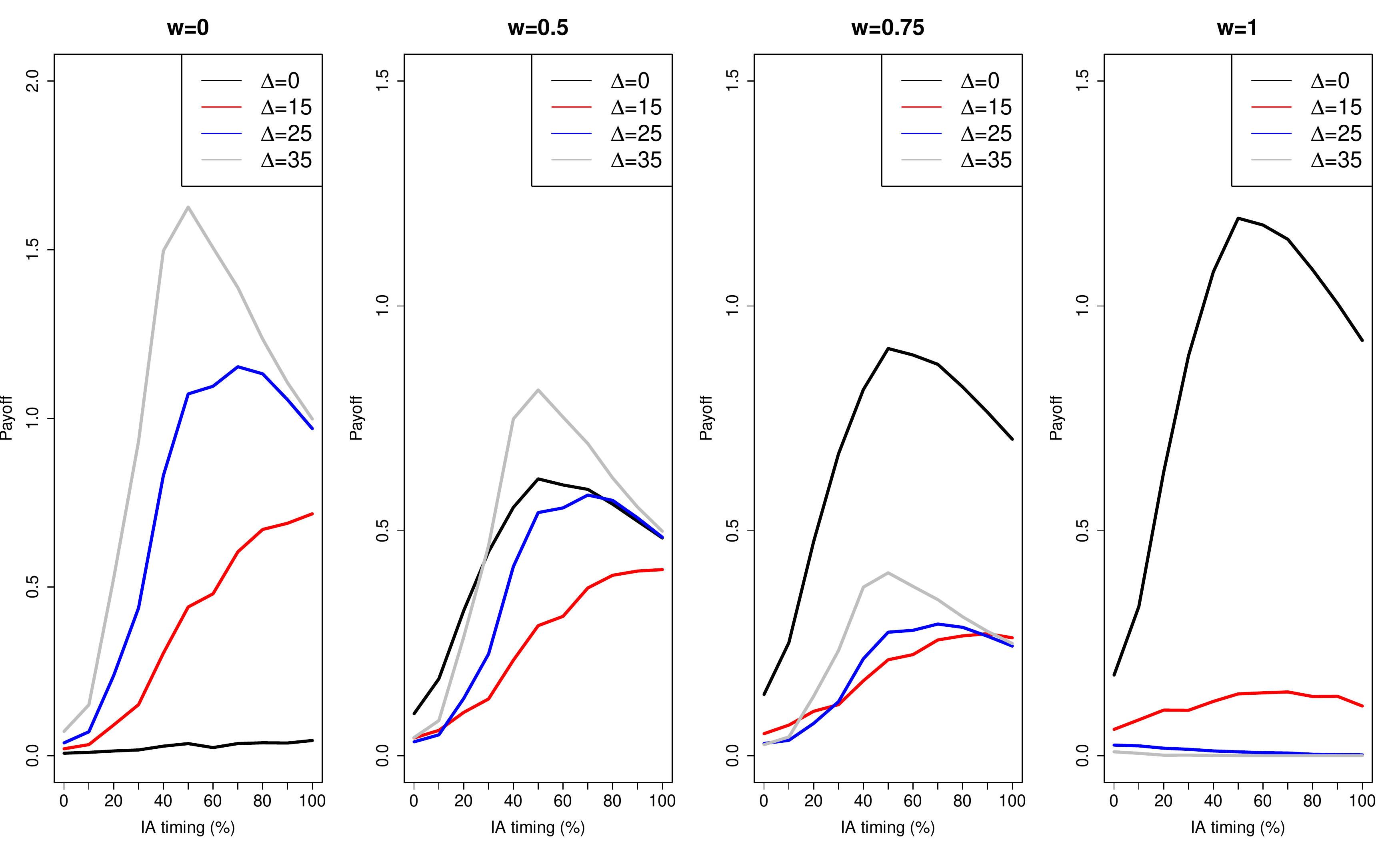}
\caption{IA timing $(\%)$ vs Full Bayesian Payoff. Each panel represents the values of payoff function with respect to different weights $w=0,0.5,0.75,1$. IA timing under different design prior with effect size: 1) $\Delta=0$, no efficacy, 2) $\Delta=15$, at minimal efficacy, 3)  $\Delta=25$, high efficacy, and 4) $\Delta=35$, surprised high efficacy. 
}
\label{payoff2}
\end{figure}

In this appendix, we include the simulation results for our fully Bayesian payoff function defined in (\ref{P1P2Bayes}).
These results are given in Figure~\ref{payoff2} and 
Table~\ref{payoff_tab2}.
Interestingly, although the shapes of payoff curves change dramatically, the optimal IA times chosen by the two payoff functions are generally comparable. 
In particular, we still see the latest optimal IA times when we expect the treatment to achieve only minimal efficacy, and the earliest optimal IA times when we expect very high efficacy. Almost flat payoff curves are observed for the weight $w=0$ under design prior corresponding to no efficacy, and for the weight $w=1$ under the design prior corresponding to high or very high efficacy. This is because under such scenarios, the payoff function only rewards the early stopping reason that is unlikely to happen (e.g., under the design prior with no efficacy, an early win would be very unusual).

\begin{table}[H]
\centering
\caption{\label{sim:results} Optimal IA timing chosen under different choices of design priors and choices of weights in the payoff}
\label{payoff_tab2}
\begin{tabular}{cccccc}
effect size & $w=0$ & $w=0.5$ & $w=0.75$ & $w=1$ \\ \hline
0    & 40 (100\%)  & 20 (50\%) & 20 (50\%)  & 20 (50\%)    \\  \hline
15   & 40 (100\%)  & 40 (100\%) & 36 (90\%)  & 28 (70\%)  \\ \hline
25   & 28 (70\%)  & 28 (70\%) & 28 (70\%)  & 0 (0\%) \\ \hline
35   & 20 (50\%) & 20 (50\%)  & 20 (50\%)  & 0 (0\%)    \\ 
   \hline 
\end{tabular} 
\end{table}

\subsection{Additional data tables}
\label{datatable}
We also include four tables (Table~\ref{sim:results1}-\ref{sim:results4}) that give detailed results from our simulation study under the four true states of nature ($\Delta=0, 15, 25$, and $35$).  Information from these tables was used in the construction of the figures in the main paper.  
\begin{table}[h]
\caption{\label{sim:results1} Results under surprised high efficacy design prior ($\Delta=35$). 
$P_1$ and $P_2$ are the marginal probabilities of early futility under null hypothesis and early efficacy under alternative hypothesis respectively. IA\_stop represents the probability of stopping the study at IA. $p_U$, $p_L$ and $p_0$ are early winner rule, final winner rule and early futility rule respectively.}
\centering
\begin{tabular}{cccccccccc}
  \hline
IA time & $P_1$ & $P_2$ & IA\_stop & Type I error & Power & $p_U$ & $p_L$ & $p_0$ & EHSS (Placebo/Treated) \\ 
  \hline
   0.00 & 0.16 & 0.03 & 0.07 & 0.051 & 0.869 & 0.998 & 0.250 & 0.975 & 21.19 / 20.75 \\ 
   4.00 & 0.26 & 0.05 & 0.14 & 0.049 & 0.869 & 0.998 & 0.250 & 0.975 & 20.78 / 20.29 \\ 
   8.00 & 0.42 & 0.13 & 0.37 & 0.051 & 0.876 & 0.998 & 0.250 & 0.975 & 20.32 / 18.94 \\ 
  12.00 & 0.56 & 0.22 & 0.56 & 0.050 & 0.879 & 0.998 & 0.250 & 0.975 & 20.01 / 17.62 \\ 
  16.00 & 0.65 & 0.40 & 0.79 & 0.048 & 0.882 & 0.996 & 0.250 & 0.975 & 19.63 / 16.26 \\ 
  20.00 & 0.74 & 0.52 & 0.90 & 0.052 & 0.885 & 0.994 & 0.250 & 0.975 & 19.31 / 15.11 \\ 
  24.00 & 0.79 & 0.60 & 0.94 & 0.048 & 0.892 & 0.994 & 0.250 & 0.975 & 18.81 / 13.83 \\ 
  28.00 & 0.84 & 0.71 & 0.98 & 0.051 & 0.899 & 0.990 & 0.250 & 0.975 & 18.14 / 12.65 \\ 
  32.00 & 0.87 & 0.79 & 0.99 & 0.051 & 0.901 & 0.986 & 0.250 & 0.975 & 17.48 / 11.63 \\ 
  36.00 & 0.90 & 0.85 & 1.00 & 0.050 & 0.903 & 0.982 & 0.250 & 0.975 & 16.76 / 10.79 \\ 
  40.00 & 0.91 & 0.90 & 1.00 & 0.050 & 0.901 & 0.976 & 0.250 & 0.975 & 15.80 / 9.86 \\ 
   \hline
\end{tabular}
\end{table}

\begin{table}[h]
\caption{\label{sim:results2} Results under high efficacy design prior ($\Delta=25$). $P_1$ and $P_2$ are the marginal probabilities of early futility under null hypothesis and early efficacy under alternative hypothesis respectively. IA\_stop represents the probability of stopping the study at IA. $p_U$, $p_L$ and $p_0$ are early winner rule, final winner rule and early futility rule respectively.}
\centering
\begin{tabular}{cccccccccc}
  \hline
IA time & $P_1$ & $P_2$ & IA\_stop & Type I error & Power & $p_U$ & $p_L$ & $p_0$ & EHSS (Placebo/Treated) \\ 
  \hline
   0.00 & 0.16 & 0.03 & 0.06 & 0.051 & 0.869 & 0.998 & 0.250 & 0.975 & 21.14 / 20.93 \\ 
   4.00 & 0.26 & 0.05 & 0.09 & 0.049 & 0.869 & 0.998 & 0.250 & 0.975 & 20.76 / 20.61 \\ 
   8.00 & 0.42 & 0.13 & 0.21 & 0.051 & 0.876 & 0.998 & 0.250 & 0.975 & 20.40 / 20.16 \\ 
  12.00 & 0.56 & 0.22 & 0.34 & 0.050 & 0.879 & 0.998 & 0.250 & 0.975 & 20.09 / 20.17 \\ 
  16.00 & 0.65 & 0.40 & 0.56 & 0.048 & 0.882 & 0.996 & 0.250 & 0.975 & 19.72 / 19.88 \\ 
  20.00 & 0.74 & 0.52 & 0.70 & 0.052 & 0.885 & 0.994 & 0.250 & 0.975 & 19.42 / 19.49 \\ 
  24.00 & 0.79 & 0.60 & 0.76 & 0.048 & 0.892 & 0.994 & 0.250 & 0.975 & 18.81 / 18.81 \\ 
  28.00 & 0.84 & 0.71 & 0.86 & 0.051 & 0.899 & 0.990 & 0.250 & 0.975 & 18.22 / 18.32 \\ 
  32.00 & 0.87 & 0.79 & 0.93 & 0.051 & 0.901 & 0.986 & 0.250 & 0.975 & 17.67 / 17.65 \\ 
  36.00 & 0.90 & 0.85 & 0.96 & 0.050 & 0.903 & 0.982 & 0.250 & 0.975 & 16.96 / 16.94 \\ 
  40.00 & 0.91 & 0.90 & 0.97 & 0.050 & 0.901 & 0.976 & 0.250 & 0.975 & 16.04 / 16.15 \\ 
   \hline
\end{tabular}
\end{table}

\begin{table}[h]
\caption{\label{sim:results3}  Results under moderate efficacy design prior ($\Delta=15$). 
$P_1$ and $P_2$ are the marginal probabilities of early futility under null hypothesis and early efficacy under alternative hypothesis respectively. IA\_stop represents the probability of stopping the study at IA. $p_U$, $p_L$ and $p_0$ are early winner rule, final winner rule and early futility rule respectively.}
\centering
\begin{tabular}{cccccccccc}
  \hline
IA time & $P_1$ & $P_2$ & IA\_stop & Type I error & Power & $p_U$ & $p_L$ & $p_0$ & EHSS (Placebo/Treated) \\ 
  \hline
   0.00 & 0.16 & 0.03 & 0.07 & 0.051 & 0.869 & 0.998 & 0.250 & 0.975 & 21.12 / 20.49 \\ 
   4.00 & 0.26 & 0.05 & 0.10 & 0.049 & 0.869 & 0.998 & 0.250 & 0.975 & 20.77 / 20.18 \\ 
   8.00 & 0.42 & 0.13 & 0.17 & 0.051 & 0.876 & 0.998 & 0.250 & 0.975 & 20.27 / 18.88 \\ 
  12.00 & 0.56 & 0.22 & 0.21 & 0.050 & 0.879 & 0.998 & 0.250 & 0.975 & 19.94 / 17.75 \\ 
  16.00 & 0.65 & 0.40 & 0.34 & 0.048 & 0.882 & 0.996 & 0.250 & 0.975 & 19.74 / 16.64 \\ 
  20.00 & 0.74 & 0.52 & 0.45 & 0.052 & 0.885 & 0.994 & 0.250 & 0.975 & 19.35 / 15.30 \\ 
  24.00 & 0.79 & 0.60 & 0.50 & 0.048 & 0.892 & 0.994 & 0.250 & 0.975 & 18.81 / 14.15 \\ 
  28.00 & 0.84 & 0.71 & 0.61 & 0.051 & 0.899 & 0.990 & 0.250 & 0.975 & 18.29 / 12.85 \\ 
  32.00 & 0.87 & 0.79 & 0.69 & 0.051 & 0.901 & 0.986 & 0.250 & 0.975 & 17.44 / 11.74 \\ 
  36.00 & 0.90 & 0.85 & 0.76 & 0.050 & 0.903 & 0.982 & 0.250 & 0.975 & 16.68 / 10.66 \\ 
  40.00 & 0.91 & 0.90 & 0.83 & 0.050 & 0.901 & 0.976 & 0.250 & 0.975 & 15.90 / 9.82 \\ 
   \hline
\end{tabular}
\end{table}

\begin{table}[h]
\caption{\label{sim:results4} Results under no efficacy design prior ($\Delta=0$). 
$P_1$ and $P_2$ are the marginal probabilities of early futility under null hypothesis and early efficacy under alternative hypothesis respectively. IA\_stop represents the probability of stopping the study at IA. $p_U$, $p_L$ and $p_0$ are early winner rule, final winner rule and early futility rule respectively.}
\centering
\begin{tabular}{cccccccccc}
  \hline
IA time & $P_1$ & $P_2$ & IA\_stop & Type I error & Power & $p_U$ & $p_L$ & $p_0$ & EHSS (Placebo/Treated) \\ 
  \hline
   0.00 & 0.16 & 0.03 & 0.16 & 0.051 & 0.869 & 0.998 & 0.250 & 0.975 & 21.03 / 19.29 \\ 
   4.00 & 0.26 & 0.05 & 0.26 & 0.049 & 0.869 & 0.998 & 0.250 & 0.975 & 20.76 / 18.30 \\ 
   8.00 & 0.42 & 0.13 & 0.43 & 0.051 & 0.876 & 0.998 & 0.250 & 0.975 & 20.25 / 13.78 \\ 
  12.00 & 0.56 & 0.22 & 0.55 & 0.050 & 0.879 & 0.998 & 0.250 & 0.975 & 19.86 / 10.11 \\ 
  16.00 & 0.65 & 0.40 & 0.66 & 0.048 & 0.882 & 0.996 & 0.250 & 0.975 & 19.61 / 7.55 \\ 
  20.00 & 0.74 & 0.52 & 0.76 & 0.052 & 0.885 & 0.994 & 0.250 & 0.975 & 19.25 / 5.73 \\ 
  24.00 & 0.79 & 0.60 & 0.81 & 0.048 & 0.892 & 0.994 & 0.250 & 0.975 & 18.59 / 4.63 \\ 
  28.00 & 0.84 & 0.71 & 0.87 & 0.051 & 0.899 & 0.990 & 0.250 & 0.975 & 18.18 / 3.91 \\ 
  32.00 & 0.87 & 0.79 & 0.91 & 0.051 & 0.901 & 0.986 & 0.250 & 0.975 & 17.42 / 3.47 \\ 
  36.00 & 0.90 & 0.85 & 0.94 & 0.050 & 0.903 & 0.982 & 0.250 & 0.975 & 16.73 / 3.08 \\ 
  40.00 & 0.91 & 0.90 & 0.97 & 0.050 & 0.901 & 0.976 & 0.250 & 0.975 & 15.85 / 2.93 \\ 
   \hline
\end{tabular}
\end{table}


\end{document}